\documentclass{quantumarticle}

\usepackage{tikz} 
\usepackage[pretty,strict,uselistings]{revquantum}
    \newrm{N}
    
    \newaffil{UWatPhysAstro}{
        Department of Physics and Astronomy,
        University of Waterloo,
        Waterloo, ON, Canada
    }
    \newaffil{MacqPhysAstro}{
        Department of Physics and Astronomy,
        Macquarie University,
        Sydney, NSW, Australia
    }
    \newaffil{UTSQSI}{
        University of Technology Sydney,
        Centre for Quantum Software and Information,
        Ultimo NSW 2007, Australia
    }    
\usepackage[sort&compress,numbers,merge]{natbib}
\usepackage[bold]{hhtensor}

\newcommand{\figurefolder}{../fig}

\newrm{ess}

\newcommand{\softwarepackagename}[1]{{#1}}

\newcommand{\qinfer}{\softwarepackagename{QInfer}}
\newcommand{\qutip}{\softwarepackagename{QuTiP}}
\newcommand{\numpy}{\softwarepackagename{NumPy}}

\newcommand{\jupyter}{\softwarepackagename{Jupyter}}
\newcommand{\ipyparallel}{\softwarepackagename{ipyparallel}}

\newcommand{\apxref}[1]{\hyperref[#1]{Appendix~\ref{#1}}}

\renewcommand{\figurefolder}{fig}
\begin{document}

\title{QInfer: Statistical inference software for quantum applications}

\author{Christopher Granade}
\email{cgranade@cgranade.com}
\homepage{www.cgranade.com}
\orcid{0000-0001-6233-3132}
\affilUSydPhys
\affilEQuSUSyd

\author{Christopher Ferrie}
\orcid{0000-0003-2736-9943}
\affilUTSQSI

\author{Ian Hincks}
\affilUWAMath
\affilIQC

\author{Steven Casagrande}
\affiliation{}

\author{Thomas Alexander}
\orcid{0000-0003-3452-0082}
\affilIQC
\affilUWatPhysAstro

\author{Jonathan Gross}
\affilCQuIC

\author{Michal Kononenko}
\affilIQC
\affilUWatPhysAstro

\author{Yuval Sanders}
\orcid{0000-0001-8003-0039}
\affilIQC
\affilUWatPhysAstro
\affilMacqPhysAstro

\date{\today}

\maketitle

\insert\footins{\footnotesize\textsf{\flushleft
    Complete source code available at DOI \href{https://dx.doi.org/10.5281/zenodo.157007}{10.5281/zenodo.157007}.
}}

\begin{abstract}
    Characterizing quantum systems through experimental data is
    critical to applications as diverse as
    metrology and quantum computing. Analyzing this experimental data
    in a robust and reproducible manner is made challenging, however,
    by the lack of readily-available software for performing principled
    statistical analysis.
    We improve the robustness and reproducibility of characterization
    by introducing an open-source library, {\textbf{\qinfer}},
    to address this need.
    Our library makes it easy to analyze data from tomography, randomized
    benchmarking, and Hamiltonian learning experiments either in post-processing,
    or online as data is acquired. \qinfer~also provides
    functionality for predicting
    the performance of proposed experimental protocols from simulated runs.
    By delivering easy-to-use characterization tools based on
    principled statistical analysis, \qinfer~helps
    address many outstanding challenges facing quantum technology.
\end{abstract}


\tableofcontents

\section{Introduction}
\label{sec:intro}

Statistical modeling and parameter estimation play a critical role in many
quantum applications. In quantum information in particular, the pursuit
of large-scale quantum information processing devices has motivated a range
of different characterization protocols, and in turn, new statistical models.
For example, quantum state and process tomography are widely used
to characterize quantum systems, and are in essence matrix-valued parameter
estimation problems \cite{dariano_quantum_2002,altepeter_ancilla-assisted_2003}.
Similarly, randomized benchmarking is now a mainstay in assessing
quantum devices, motivating the use of rigorous statistical analysis
\cite{wallman_randomized_2014} and algorithms
\cite{granade_accelerated_2015}. Quantum metrology, meanwhile, is intimately
concerned with what parameters can be extracted from measurements of
a physical system, immediately necessitating a statistical view
\cite{holevo_statistical_2001,helstrom_quantum_1976}.

The prevalence of statistical modeling in quantum applications should not be
surprising: quantum mechanics is an inherently statistical theory, thus
inference is an integral part of both experimental and theoretical practice.
In the former, experimentalists need to model their systems and infer
the value of parameters for the purpose of improving control as well as
validating performance. In the latter, numerical experiments utilizing
simulated data are now commonplace in theoretical studies, such that the same
inference problems are encountered usually as a necessary step to answer
questions about optimal data processing protocols or experiment design.
In both cases, we lack tools to rapidly prototype and access inference
strategies; {\textbf{\qinfer}}~addresses this need by providing a modular
interface to a Monte Carlo algorithm for performing statistical inference.

Critically, in doing so, \qinfer~also supports and enables
open and reproducible research practices.  Parallel to the challenges faced in
many other disciplines \cite{goldacre_scientists_2015}, physics research cannot
long survive its own current practices.  Open access, open source, and
open data provide an indispensable means for research to be reproducible,
ensuring that research work is useful to the communities invested in that
research \cite{stodden_best_2014}.  In the particular context of quantum
information research, open methods are especially critical given the impact of
statistical errors that can undermine the claims of published research
\cite{ioannidis_why_2005,hoekstra_robust_2014}.
Ensuring the reproducibility of research is critical for evaluating
the extent to which statistical and methodological errors undermine the
credibility of published research \cite{ioannidis_how_2014}.

\qinfer~also constitutes an important step towards a more general framework
for quantum verification and validation (QCVV).
As quantum information processor prototypes become more complex, the challenge of ensuring that
noise processes affecting these devices conform to some agreed-upon standard becomes more difficult.
This challenge can be managed, at least in principle, by developing confidence in the truth
of certain simplifying assumptions and approximations.  The value of randomized benchmarking, for example,
depends strongly upon the extent to which noise is approximately Pauli \cite{sanders_bounding_2016}.
\qinfer~provides a valuable framework for the design of automated and efficient noise assessment methods
that will enable the comparison of actual device performance to the specifications demanded by theory.

To the end of enabling reproducible and accessible research,
and hence providing a reliable process for interpreting advances
in quantum information processing, we base \qinfer~using
openly-available tools such as the Python programming language,
the IPython interpreter, and
\jupyter~\cite{perez_ipython:_2007,jupyter_development_team_jupyter_2016}.
Jupyter in particular has already proven to be an invaluable tool
for reproducible research, in that it provides a powerful framework
for describing and explaining research software
\cite{piccolo_tools_2016,*vries_using_2015,*donoho_reproducible_2015}.
We provide our library under an open-source license
along with examples \cite{qinfer-examples}~of how to use
\qinfer~to support reproducible research practices.
In this way, our library builds on and supports recent efforts to develop
reproducible methods for physics research \cite{dolfi_model_2014}.

\qinfer~is a mature open-source software library written in the Python
programming language which has now been  extensively tested in a wide range of
inferential problems by various research groups. Recognizing its maturity
through its continuing development, we now formally release version 1.0. This
maturity has given its developers the opportunity to step back and focus on
the accessibility of \qinfer~such that other researchers can benefit from its
utility. This short paper is the culmination of that effort. A full Users'
Guide is available in the ancillary files.

We proceed as following. In \autoref{sec:bayes}, we give a brief introduction to Bayesian inference
and particle filtering, the numerical algorithm we use to implement Bayesian
updates. In \autoref{sec:quantum-models}, we describe applications of
\qinfer~to common tasks in quantum information processing. Next, we describe
in \autoref{sec:addl-functionality} additional features of \qinfer~before
concluding in \autoref{sec:conclusions}.

\section{Inference and Particle Filtering}
\label{sec:bayes}

\qinfer~is primarily intended to serve as a toolset for implementing
Bayesian approaches to statistical inference.  In this section,
we provide a brief review of the Bayesian formalism for statistical inference.
This section is not intended to be comprehensive; our aim is rather to establish
the language needed to describe the \qinfer~codebase.

In the Bayesian paradigm, statistical inference is the process of evaluating
data obtained by sampling an unknown member of a family of related probability
distributions, then using these samples to assign a relative plausibility to each
distribution. Colloquially, we think of this family of distributions as a \emph{model}
parameterized by a vector $\vec{x}$ of model parameters.
We then express the probability that a dataset $D$ was obtained from the model
parameters $\vec{x}$ as $\Pr(D|\vec{x})$ and read it as ``the probability of $D$ \emph{given} that
the model specified by $\vec{x}$ is the correct model.''  The function
$\Pr(\cdot|\vec{x})$ is called the likelihood function, and computing it is
equivalent to \emph{simulating} an experiment%
\footnote{Here, we use the word ``simulation'' in the sense
of what \citet{van_den_nest_simulating_2009}
terms ``strong simulation,'' as opposed to drawing data consistent
with a given model (``weak simulation'').}. For example, the Born rule is a likelihood
function, in that it maps a known or hypothetical quantum density matrix $\vec{x} \equiv \rho$ to a
distribution over measurement outcomes of a measurement $D \in \{E, \id - E\}$ via
\begin{equation}
    \Pr(D=E|\vec{x}) = \operatorname{Tr}(E \rho).
\end{equation}

The problem of estimating model parameters is as follows.
Suppose an agent is provided with a dataset $D$ and is tasked with
judging the probability that the model specified by a given vector $\vec{x}$
is in fact the correct one.  According to Bayes' rule,
\begin{equation}
    \Pr(\vec{x}|D) =\frac{\Pr(D|\vec{x})\Pr(\vec{x})}{\Pr(D)},
\end{equation}
where $\Pr(\vec{x})$ is a probability distribution called the \emph{prior}
distribution and $\Pr(\vec{x}|D)$ is called the \emph{posterior} distribution.
If the agent is provided with a prior distribution, then they can estimate
parameters using Bayes' rule. Note that $\Pr(D)$ can be computed through marginalization,
which is to say that the value can in principle be calculated via the equation
\begin{equation}
    \Pr(D) = \int_{\vec{x}} \Pr(D|\vec{x}) \Pr(\vec{x}) d\vec{x}.
\end{equation}
For the inference algorithm used by \qinfer,
$\Pr(D)$ is an easily computed normalization constant and
there is no need to compute a possibly complicated integral.

Importantly, we will demand that the agent's data processing approach works in an
\emph{iterative} manner.  Consider the example in which the data $D$ is in fact
a set $D = \{d_1, \dots, d_N\}$ of individual observations.
In most if not all classical applications,
each individual datum is distributed independently of the
rest of the data set, conditioned on the true state.
Formally, we write that for all $j$ and $k$ such that
$j\ne k$, $d_{j} \perp d_{k} \mid \vec{x}$.  This may not hold in quantum models
where measurement back-action can alter the state.
In such cases, we can simply redefine what the parameters $\vec{x}$ label,
such that this independence property can be taken as a convention,
instead of as an assumption.  Then, we have that
\begin{equation}
    \Pr(\vec{x}|d_1,\ldots,d_N) = \frac{\Pr(d_N|\vec{x})\Pr(\vec{x}|d_1,\ldots,d_{N-1})}{\Pr(d_N)}.
\end{equation}
In other words, the agent can process the data \emph{sequentially}
where the prior for each successive datum is the posterior from the last.

This Bayes update can be solved analytically in some important special cases,
such as frequency estimation \cite{ferrie_how_2013,sergeevich_characterization_2011},
but is more generally intractable.  Thus, to develop a robust and generically
useful framework for parameter estimation, the agent relies on numerical algorithms.
In particular, \qinfer~is largely concerned with the \emph{particle filtering}
algorithm \cite{doucet_tutorial_2011}, also known as the sequential Monte Carlo
(SMC) algorithm.  In the context of quantum information, SMC was first proposed
for learning from continuous measurement records \cite{chase_single-shot_2009},
and has since been used to learn from state tomography \cite{huszar_adaptive_2012},
Hamiltonian learning \cite{granade_robust_2012}, and randomized benchmarking
\cite{granade_accelerated_2015}, as well as other applications.

The aim of particle filtering is to replace a continuous probability distribution
$\Pr(\vec{x})$ with a discrete approximation
\begin{equation}
    \label{eq:smc-approx}
    \sum_k w_k \delta(\vec{x}-\vec{x}_k),
\end{equation}
where $\vec{w} = (w_k)$ is a vector of probabilities.  The entry $w_k$ is called
the \emph{weight} of the \emph{particle}, labeled $k$, and $\vec{x}_k$ is
the \emph{location} of particle $k$.

Of course, the particle filter $\sum_k w_k \delta(\vec{x} - \vec{x}_k)$
does not directly approximate $\Pr(\vec{x})$ as a distribution;
the particle filter, if considered as a distribution, is supported
only a discrete set of points.  Instead, the particle filter is used to approximate
expectation values: if $f$ is a function whose domain is the set of
model vectors $\vec{x}$, we want the particle filter to satisfy
\begin{equation}
    \int f(\vec{x}) \Pr(\vec{x}) d\vec{x} \approx \sum_k w_k f(\vec{x}_k).
\end{equation}

The posterior distribution can also be approximated using a particle filter.
In fact, a posterior particle filter can be computed directly from
a particle filter for the prior distribution as follows.
Let $\{(w_k, \vec{x}_k)\}$ be the set of weights and locations for
a particle filter for some prior distribution $\Pr(\vec{x})$.
We then compute a particle filter $\{(w_k^\prime, \vec{x}_k^\prime)\}$ for
the posterior distribution by setting $\vec{x}_k^\prime = \vec{x}_k$ and
\begin{equation}
    \label{eq:smc-update}
    w_k^\prime = \frac{w_k \Pr(D|\vec{x}_k)}{\sum_j w_j \Pr(D|\vec{x}_j)},
\end{equation}
where $D$ is the data set used in the Bayesian update.
In practice, updating the weights in this fashion causes the particle filter
to become unstable as data is collected; by default, \qinfer~will
periodically apply the Liu--West algorithm to restore stability
\cite{liu_combined_2001}.  See Appendix~\ref{apx:resampling} for details.

At any point during the processing of data,
the expectation of any function with respect to the posterior is approximated as
\begin{equation}
    \expect[f(\vec{x}) | D] \approx \sum_k w_k(D) f(\vec{x}_k).
\end{equation}
In particular, the expected error in $\vec{x}$ is given by the posterior
covariance, $\Cov(\vec{x} | D) \defeq \expect[\vec{x} \vec{x}^\T | D] -
\expect[\vec{x} | D] \expect^\T[\vec{x} | D]$.
This can be used, for instance, to adaptively choose experiments which minimize
the posterior variance \cite{granade_robust_2012}. This approach has been used
to exponentially improve the number of samples required in frequency estimation
problems \cite{sergeevich_characterization_2011,ferrie_how_2013}, and in
phase estimation \cite{berry_optimal_2000,higgins_entanglement-free_2007}.
Alternatively, other cost functions can be considered, such as
the information gain \cite{huszar_adaptive_2012,struchalin_experimental_2016}.
\qinfer~allows for quickly computing either the expected posterior variance or
the information gain for proposed experiments, making it straightforward
to develop adaptive experiment design protocols.

The functionality exposed by \qinfer~follows a simple object model, in which
the experiment is described in terms of a \emph{model}, and background
information is described in terms of a \emph{prior distribution}. Each of
these classes is \emph{abstract}, meaning that they define what behavior a
\qinfer~user must specify in order to fully specify an inference procedure.
For convenience, \qinfer~provides several pre-defined implementations of each,
as we will see in the following examples.
Concrete implementations of a model and a prior distribution are then used
with SMC to \emph{update} the prior based on data. In summary, the
iterative approach described above is formalized in terms of the following Python object model:

\begin{description}
    \item[\emph{class} qinfer.Distribution:]
        \leavevmode 
        \begin{description}
            \item[\emph{abstract} sample($n$):]
                Returns $n$ samples from the represented distribution.
        \end{description}
    \item[\emph{class} qinfer.Model:]
        \leavevmode
        \begin{description}
            \item[\emph{abstract} likelihood($d$, $\vec{x}$, $e$):]
                Returns an evaluation of the likelihood
                function $\Pr(d | \vec{x}; e)$ for a single datum $d$,
                a vector of model parameters $\vec{x}$ and an experiment $e$.
            \item[\emph{abstract} are\_models\_valid($\vec{x}$):]
                Evaluates whether $\vec{x}$ is a valid assignment of model
                parameters.
        \end{description}
    \item[\emph{class} qinfer.SMCUpdater:]
        \leavevmode
        \begin{description}
            \item[update($d$, $e$):]
                Computes the Bayes update \autoref{eq:smc-update}
                for a single datum (that is, $D = \{d\}$).
            \item[est\_mean():]
                Returns the current estimate $\hat{\vec{x}} = \expect[\vec{x}]$.
        \end{description}
\end{description}

A complete description of the \qinfer~object model can be found
in the Users' Guide.
Notably \lstinline+qinfer.SMCUpdater+ relies only on the behavior specified by
each of the abstract classes in this object model.
Thus, it is straightforward for
the user to specify their own prior and likelihood
function by either implementing these classes (as in the
example of \apxref{apx:custom-model}), or by using one
of the many concrete implementations provided with \qinfer.

The concrete implementations provided with \qinfer~are useful in a
range of common applications, as described in the next Section.
We will demonstrate how these classes
are used in practice with examples drawn from quantum information applications.
We will also consider the \lstinline+qinfer.Heuristic+ class, which is useful
in contexts such as online adaptive experiments and simulated experiments.

\section{Applications in Quantum Information}
\label{sec:quantum-models}

In this Section, we describe various possible applications of \qinfer~to
existing experimental protocols.
In doing so, we highlight both functionality built-in to
\qinfer~and how this functionality can be readily extended
with custom models and distributions.
We begin with the problems of phase and frequency learning,
then describe the use of \qinfer~for state and process tomography,
and conclude with applications to randomized benchmarking.

\subsection{Phase and Frequency Learning}
\label{sec:hamiltonian-learning}

One of the primary applications for particle filtering
is for learning the Hamiltonian $H$ under which a quantum system evolves
\cite{granade_robust_2012}.
For instance, consider the single-qubit
Hamiltonian $H = \omega \sigma_z / 2$ for an unknown parameter $\omega$. An
experiment on this qubit may then consist of preparing a state $\ket{+} =
(\ket{0} + \ket{1}) / \sqrt{2}$, evolving for a time $t$ and then measuring
in the $\sigma_x$ basis. This model commonly arises from Ramsey interferometry,
and gives a likelihood function
\begin{gather}
    \label{eq:simple-prec-like}
    \begin{aligned}
        \Pr(0 | \omega; t) & = \big| \bra{+} \e^{-\ii \omega t \sigma_z / 2} \ket{+} \big|^2 \\
                           & = \cos^2(\omega t / 2).
    \end{aligned}
\end{gather}
Note that this is also the same model for Rabi interferometry as well, with the
interpretation of $H$ as drive term rather than the internal Hamiltonian for
a system. Similarly, this model forms the basis of Bayesian and maximum
likelihood approaches to phase estimation.

In any case, \qinfer~implements \autoref{eq:simple-prec-like} as the
\lstinline+SimplePrecessionModel+ class, making it easy to quickly perform
Bayesian inference for Ramsey or Rabi estimation problems. We demonstrate
this in \autoref{lst:freq-est-updater-loop}, using \lstinline+ExpSparseHeuristic+ to select
the $k$th measurement time $t_k = (9 / 8)^k$, as suggested by analytic
arguments \cite{ferrie_how_2013}.

\onecolumngrid
\begin{lstlisting}[
        caption={Frequency estimation example using \lstinline+SimplePrecessionModel+.},
        label={lst:freq-est-updater-loop},
        emph={SimplePrecessionModel}
    ]
    >>> from qinfer import *
    >>> model = SimplePrecessionModel()
    >>> prior = UniformDistribution([0, 1])
    >>> n_particles = 2000
    >>> n_experiments = 100
    >>> updater = SMCUpdater(model, n_particles, prior)
    >>> heuristic = ExpSparseHeuristic(updater)
    >>> true_params = prior.sample()
    >>> for idx_experiment in range(n_experiments):
    ...     experiment = heuristic()
    ...     datum = model.simulate_experiment(true_params, experiment)
    ...     updater.update(datum, experiment)
    >>> print(updater.est_mean())
\end{lstlisting}

\begin{figure}
    \begin{center}
        \includegraphics[width=0.6\columnwidth]{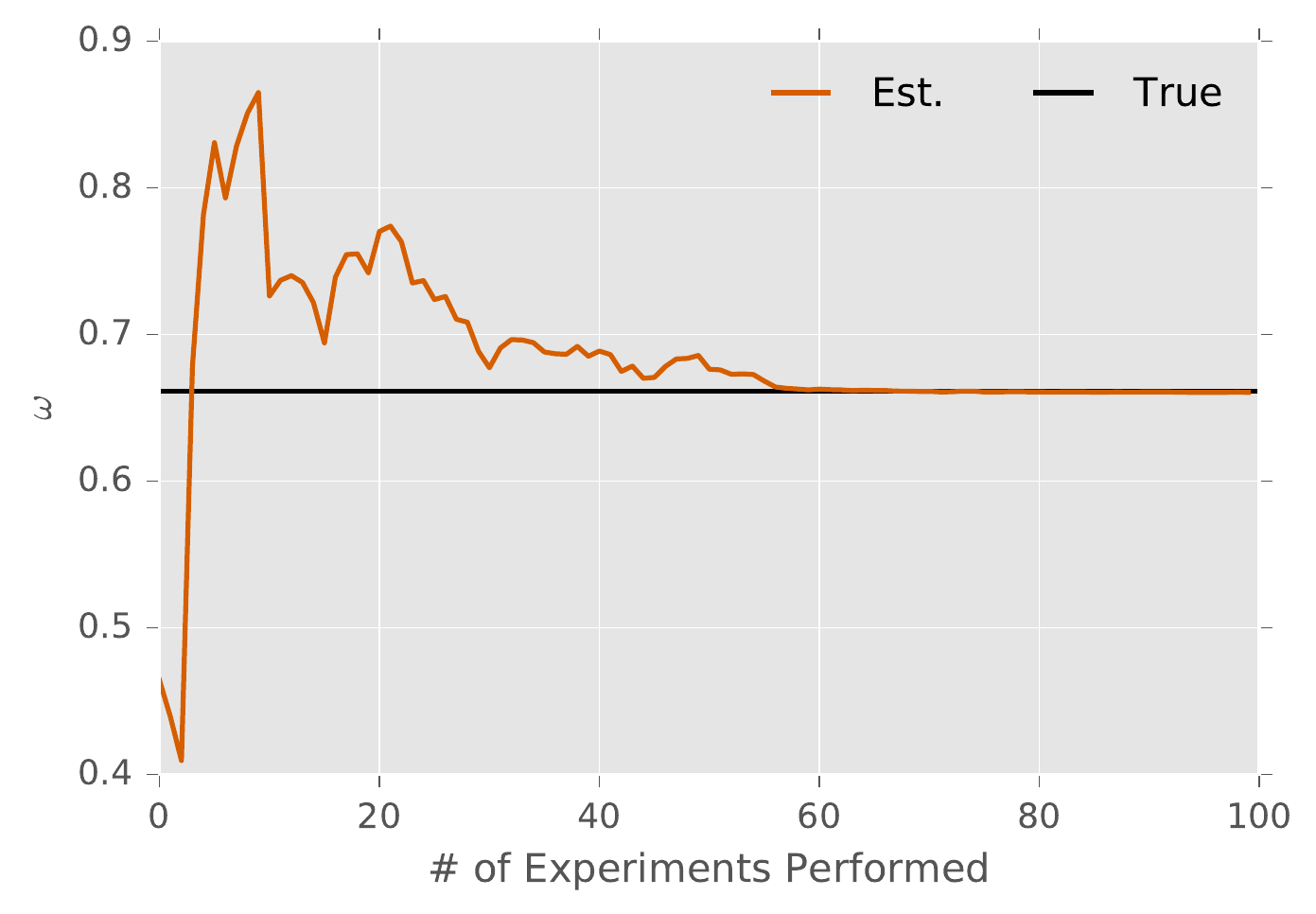}
        \caption{\label{fig:freq-est-updater-loop}
            Frequency estimate obtained using \autoref{lst:freq-est-updater-loop}
            as a function of the number of experiments performed.
        }
    \end{center}
\end{figure}
\twocolumngrid

More complicated models for learning Hamiltonians with particle
filtering have also been
considered \cite{wiebe_hamiltonian_2014,wiebe_quantum_2014-1,wiebe_quantum_2015,stenberg_efficient_2014};
these can be readily implemented in \qinfer~as custom
models by deriving from the \lstinline+Model+ class,
as described in \apxref{apx:custom-model}.

\subsection{State and Process Tomography}
\label{sec:tomography}

Though originally conceived of as a algebraic inverse problem,
quantum tomography is also a problem of parameter estimation.
Many have also considered the problem in a Bayesian framework
\cite{jones_principles_1991,blume-kohout_optimal_2010} and
the sequential Monte Carlo algorithm has been used in both
theoretical and experimental studies
\cite{ferrie_quantum_2014,granade_practical_2016,huszar_adaptive_2012,
kravtsov_experimental_2013,struchalin_experimental_2016}.

To define the model, we start with a basis for traceless Hermitian operators
$\{B_j\}_{j=1}^{d^2-1}$.  In the case of a qubit, this could be the basis of
Pauli matrices, for example.  Then, any state $\rho$ can be written
\begin{equation}
    \rho = \frac{\id}{d}+\sum_{j=1}^{d^2-1} \theta_j B_j,
\end{equation}
for some vector of parameters $\vec\theta$.
These parameters must be constrained such that $\rho\geq0$.

In the simplest case, we can consider two-outcome measurements represented
by the pair $\{E,\id-E\}$.  The Born rule defines the likelihood function
\begin{equation}
     \Pr(E|\rho)=\Tr(\rho E).
 \end{equation}
For multiple measurements, we simply iterate.  For many trials of the same
measurement, we can use a \emph{derived model} as discussed below.

\qinfer's \lstinline+TomographyModel+ abstracts many of the implementation
details of this problem, exposing tomographic models and estimates in terms
of \qutip's \lstinline+Qobj+ class \cite{pitchford_qutip_2015}.
This allows for readily integrating
\qinfer~functionality with that of \qutip, such as fidelity metrics,
diamond norm calculation, and other such manipulations.

Tomography support in \qinfer~requires one of the bases mentioned above in
order to parameterize the state. Many common choices of basis are included as
\lstinline+TomographyBasis+ objects, such as the Pauli or Gell-Mann bases.
Many of the most commonly used priors are already implemented as a
\qinfer~\lstinline+Distribution+.

\onecolumngrid
\begin{lstlisting}[
        caption={Rebit state tomography example using \lstinline+TomographyModel+.},
        label={lst:rebit-tomo},
        emph={TomographyModel}
    ]
    >>> from qinfer import *
    >>> from qinfer.tomography import *
    >>> basis = pauli_basis(1) # Single-qubit Pauli basis.
    >>> model = TomographyModel(basis)
    >>> prior = GinibreReditDistribution(basis)
    >>> updater = SMCUpdater(model, 8000, prior)
    >>> heuristic = RandomPauliHeuristic(updater)
    >>> true_state = prior.sample()
    >>>
    >>> for idx_experiment in range(500):
    >>>     experiment = heuristic()
    >>>     datum = model.simulate_experiment(true_state, experiment)
    >>>     updater.update(datum, experiment)
\end{lstlisting}

\begin{figure}
    \begin{center}
        \includegraphics[width=0.7\columnwidth]{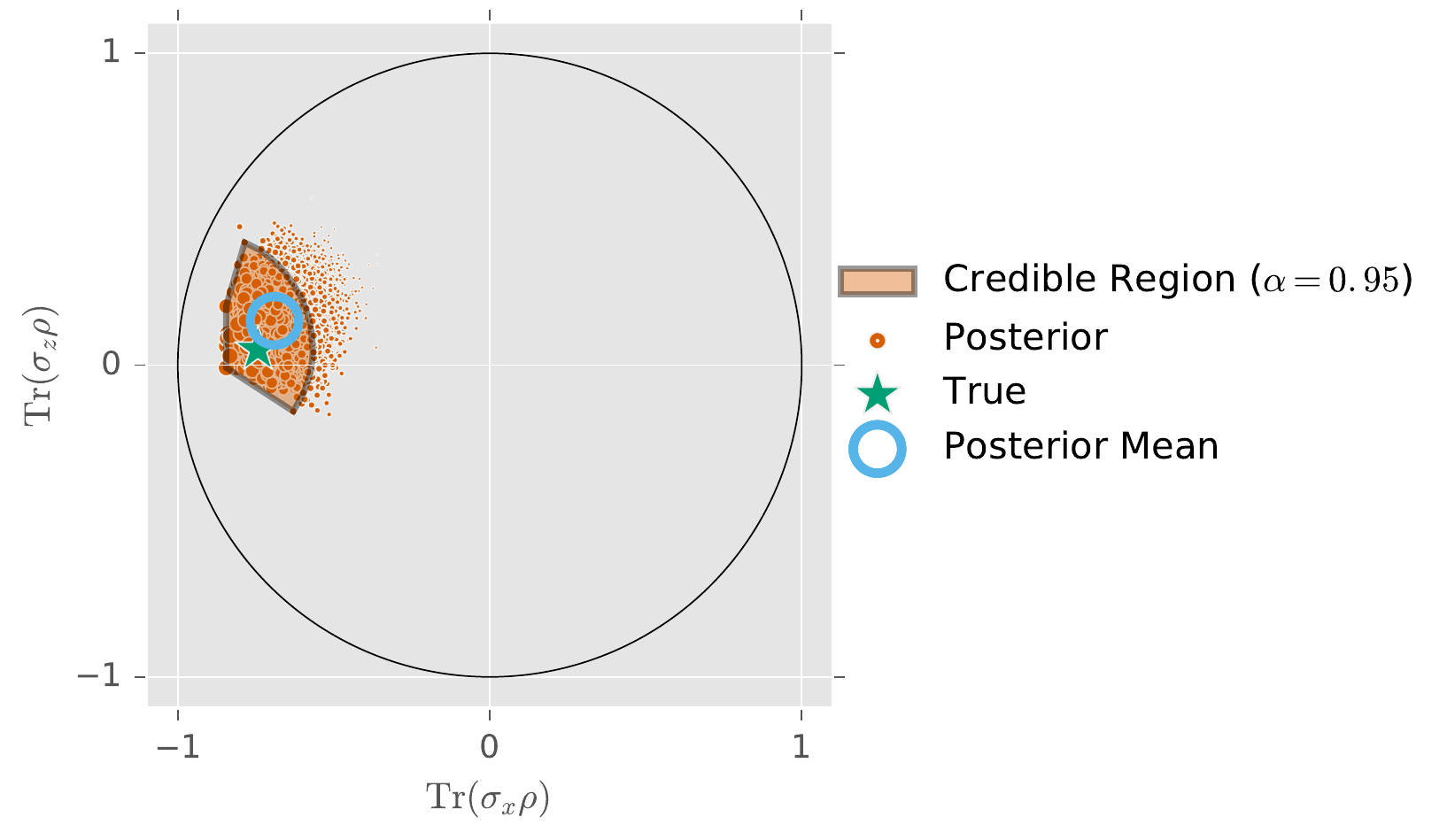}
        \caption{\label{fig:rebit-tomo}
            Posterior over rebit states after 100 random Pauli measurements,
            each repeated five times, as implemented by \autoref{lst:rebit-tomo}.
        }
    \end{center}
\end{figure}
\twocolumngrid

For simulations, common randomized measurement choices are already
implemented. For example, \lstinline+RandomPauliHeuristic+ chooses random
Pauli measurements for qubit tomography.

In \autoref{lst:rebit-tomo}, we demonstrate \qinfer's tomography support for a
\emph{rebit}. By analogy to the Bloch sphere, a rebit may be represented by a
point in the unit disk, making rebit tomography useful for plotting examples.
More generally, with different choices of basis, \qinfer~can be used for
qubits or higher-dimensional states. For example, recent work has demonstrated
the use of \qinfer~for tomography procedures on seven-dimensional states
\cite{granade_paqt_2016}. Critically, \qinfer~provides a \emph{region estimate}
for this example, describing a region that has a 95\% probability of containing
the true state. We will explore region estimation further in \autoref{sec:regions}.

Finally, we note that process tomography is a special case of state tomography
\cite{granade_practical_2016}, such that the same functionality described
above can also be used to analyze process tomography experiments. In
particular, the \lstinline+qinfer.ProcessTomographyHeuristic+ class represents
the experiment design constraints imposed by process tomography, while
\lstinline+BCSZChoiDistribution+ uses the distribution over completely
positive trace-preserving maps proposed by \citet{bruzda_random_2009} to
represent a prior distribution over the Choi states of random channels.

\subsection{Randomized Benchmarking}
\label{sec:rb}

In recent years, randomized benchmarking (RB) has reached a critical role in
evaluating candidate quantum information processing systems. By
using random sequences of gates drawn from the Clifford group, RB provides
a likelihood function that depends on the fidelity with which each Clifford
group element is implemented, allowing for estimates of that fidelity to
be drawn from experimental data \cite{granade_accelerated_2015}.

In particular, suppose that each gate is implemented with fidelity $F$, and
consider a fixed initial state and measurement. Then, the survival
probability over sequences of length $m$ is given by
\cite{magesan_scalable_2011}
\begin{equation}
    \label{eq:rb-like}
    \Pr(\text{survival} | p, A, B; m) = A p^m + B,
\end{equation}
where $p \defeq (d F - 1) / (d - 1)$, $d$ is the dimension of the system
under consideration, and where $A$ and $B$ describe the state preparation
and measurement (SPAM) errors. Learning the model $\vec{x} = (p, A, B)$
thus provides an estimate of the fidelity of interest $F$.

The likelihood function for randomized benchmarking is extremely
simple, and requires only scalar arithmetic to compute, making it especially
useful for avoiding the computational overhead typically required to characterize
large quantum systems with classical resources. Multiple generalizations of RB
have been recently developed which extend these benefits to estimating
crosstalk \cite{gambetta_characterization_2012}, coherence \cite{wallman_estimating_2015},
and to estimating fidelities
of non-Clifford gates \cite{cross_scalable_2016,harper_estimating_2016}.
RB has also been extended to provide
tomographic information as well \cite{kimmel_robust_2014}.
The estimates provided by randomized benchmarking have also been applied
to design improved control sequences \cite{ferrie_robust_2015,egger_adaptive_2014}.

\qinfer~supports RB experiments through the
\lstinline+qinfer.RandomizedBenchmarkingModel+ class. For common priors,
\qinfer~also provides a simplified interface, \lstinline+qinfer.simple_est_rb+,
that reports the mean and covariance over an RB model given experimental
data. We provide an example in \autoref{lst:rb-simple-est}.

\onecolumngrid
\begin{lstlisting}[
        caption={Randomized benchmarking example using \lstinline+simple_est_rb+.},
        label={lst:rb-simple-est},
        emph={simple_est_rb}
    ]
    >>> from qinfer import *
    >>> import numpy as np
    >>> p, A, B = 0.95, 0.5, 0.5
    >>> ms = np.linspace(1, 800, 201).astype(int)
    >>> signal = A * p ** ms + B
    >>> n_shots = 25
    >>> counts = np.random.binomial(p=signal, n=n_shots)
    >>> data = np.column_stack([counts, ms, n_shots * np.ones_like(counts)])
    >>> mean, cov = simple_est_rb(data, n_particles=12000, p_min=0.8)
    >>> print(mean, np.sqrt(np.diag(cov)))
\end{lstlisting}

\begin{figure}
    \begin{center}
        \includegraphics[width=0.95\columnwidth]{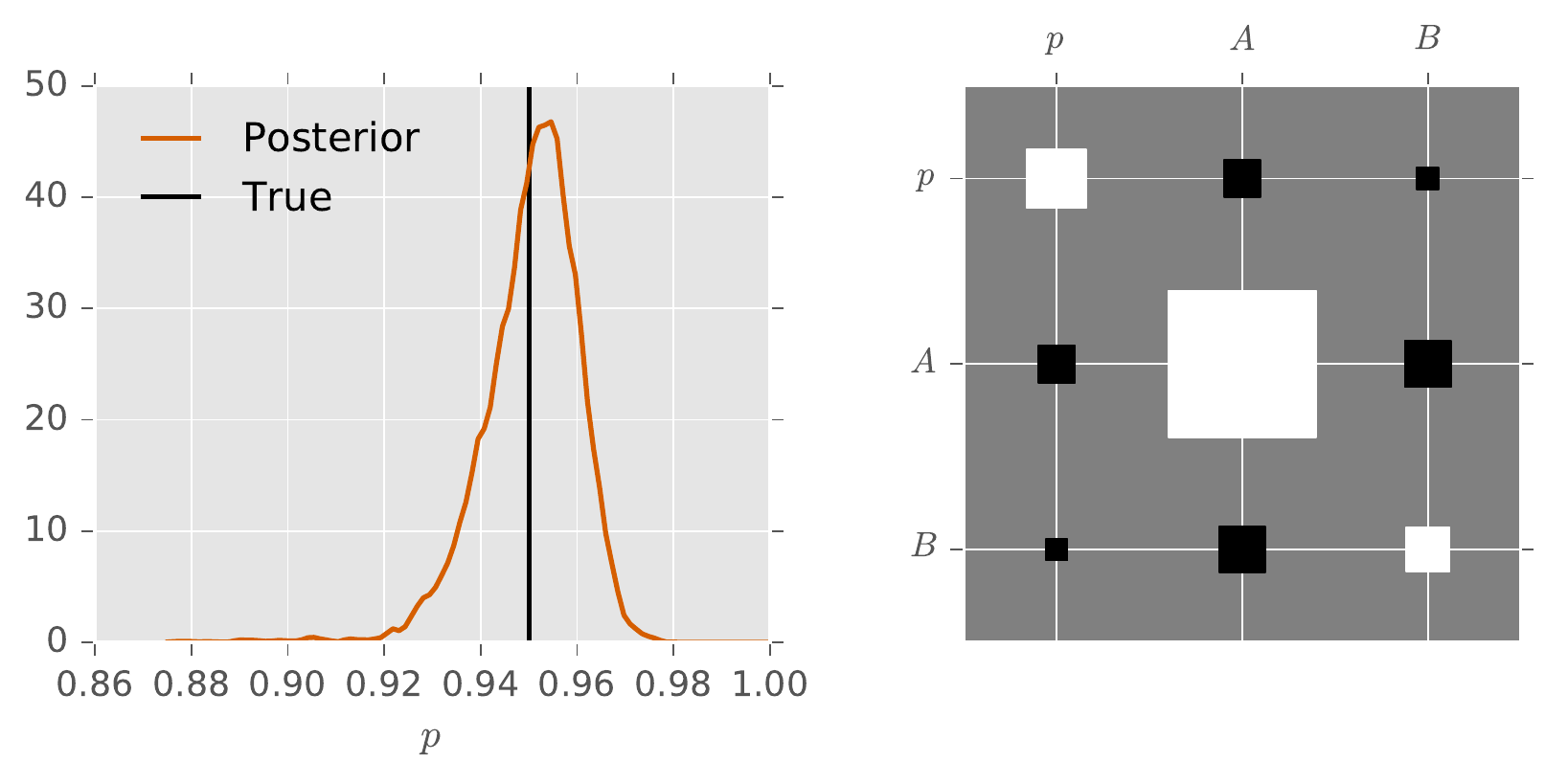}
        \caption{\label{fig:rb-simple-est}
            (Left) Posterior over the randomized benchmarking decay rate
            parameter $p$ after measuring 25 sequences at each of 201
            sequence lengths, as described in \autoref{lst:rb-simple-est}.
            (Right) The posterior covariance matrix over all three randomized
            benchmarking parameters $\vec{x} = (p, A, B)$, represented as a
            Hinton diagram. White squares indicate positive elements, while
            black squares indicate negative elements, and the relative sizes
            indicate magnitude of each element.
        }
    \end{center}
\end{figure}
\twocolumngrid

\section{Additional Functionality}
\label{sec:addl-functionality}

Having introduced common applications for \qinfer, in this Section we describe
additional functionality which can be used with each of these applications, or
with custom models.

\subsection{Region Estimation and Error Bars}
\label{sec:regions}

As an alternative to specifying the entire posterior distribution approximated
by \lstinline+qinfer.SMCUpdater+, we provide methods for reporting credible
regions over the posterior, based on covariance ellipsoids, convex hulls, and
minimum volume enclosing ellipsoids \cite{ferrie_quantum_2014}. These region
estimators provide a rigorous way of summarizing one's uncertainty following
an experiment (colloquially referred to as ``error bars''), and owing to the
Bayesian approach, do so in a manner consistent with experimental experience.

Posterior credible regions can be found by using the \lstinline+SMCUdater.est_credible_region+ method. This method returns a set of particles such that the sum of their weights corresponding weights is at least a specified ratio of the total weight. For example, a 95\% credible regions is represented as a collection of particles whose weight sums to at least 0.95.

This does not necessarily admit a very compact description since many of the particles would be interior to the regions. In such cases, it is useful to find region estimators containing all of the particles describing a credible region. The \lstinline+SMCUpdater.region_est_hull+ method does this by finding a convex hull of the credible particles. Such a hull is depicted in \autoref{fig:rebit-tomo}.

The convex hull of an otherwise random set of points is also not necessarily easy to describe or intuit. In such cases, \lstinline+SMCUdpater.region_est_ellipsoid+ finds the minimum-volume enclosing ellipse (MVEE) of the convex hull region estimator. As the name suggests, this is the smallest ellipsoid containing the credible particles. It is strictly larger than the hull and thus maintains credibility. Ellipsoids are specified by their center and covariance matrix. Visualizing the covariance matrix can also usually provide important diagnostic information, as in \autoref{fig:rb-simple-est}. In that example, we can quickly see that the $p$ and $A$ parameters estimated from a randomized benchmarking experiment are anti-correlated, such that we can explain more preparation and measurement errors by assuming better gates and vice versa.

\subsection{Derived Models}
\label{sec:derived}

\qinfer~allows for the notion of a \textit{model chain}, where the likelihood
of a given model in the chain is a function of the likelihoods of models
below it, and possibly new model or experiment parameters.
This abstraction is useful for a couple of reasons.
It encourages more robust programs, since models in the chain will
often be debugged independently.
It also often makes writing new models easier since part of the chain
may be included by default in the \qinfer~library,
or may overlap with other similar models the user is implementing.
Finally, in quantum systems, it is common to have a likelihood
function which is most naturally expressed as a hierarchical
probability distribution, with base models describing quantum physics,
and overlying models describing measurement processes.

Model chains are typically implemented through the use of the abstract class
\lstinline+DerivedModel+.
Since this class itself inherits from the \lstinline+Model+ class,
subclass instances must provide standard model properties and methods such as
\lstinline+likelihood+, \lstinline+n_outcomes+, and \lstinline+modelparam_names+.
Additionally, \lstinline+DerivedModel+ accepts an argument \lstinline+model+,
referring to the \textit{underlying model} directly below it in the model chain.
Class properties exist for referencing models at arbitrary depths in the chain,
all the way down to the \textit{base model}.

As an example, consider a base model which is the precession model discussed
in \autoref{sec:hamiltonian-learning}.
This is a two-outcome model whose outcomes correspond to measuring the state
$\ket{+}$ or the orthogonal state $\ket{-}$, which can be viewed as flipping
a biased coin.
Perhaps an actual experiment of this system consists of flipping the coin $N$
times with identical settings, where the individual results are not
recorded, only the total number $n_+$ of $\ket{+}$ results.
In this case, we can concatenate this base model with the built-in
\lstinline+DerivedModel+ called \lstinline+BinomialModel+.
This model adds an additional experiment parameter \lstinline+n_meas+
specifying how many times the underlying model's coin
is flipped in a single experiment.

\onecolumngrid
\begin{lstlisting}[
        emph={BinomialModel},
        caption={Frequency estimation with the derived \lstinline+BinomialModel+
                 and a linear time-sampling heuristic.},
        label={lst:derived-model-updater-loop}
    ]
    >>> from qinfer import *
    >>> import numpy as np
    >>> model = BinomialModel(SimplePrecessionModel())
    >>> n_meas = 25
    >>> prior = UniformDistribution([0, 1])
    >>> updater = SMCUpdater(model, 2000, prior)
    >>> true_params = prior.sample()
    >>> for t in np.linspace(0.1,20,20):
    ...     experiment = np.array([(t, n_meas)], dtype=model.expparams_dtype)
    ...     datum = model.simulate_experiment(true_params, experiment)
    ...     updater.update(datum, experiment)
    >>> print(updater.est_mean())
\end{lstlisting}

\begin{figure}
    \begin{center}
        \includegraphics[width=0.7\columnwidth]{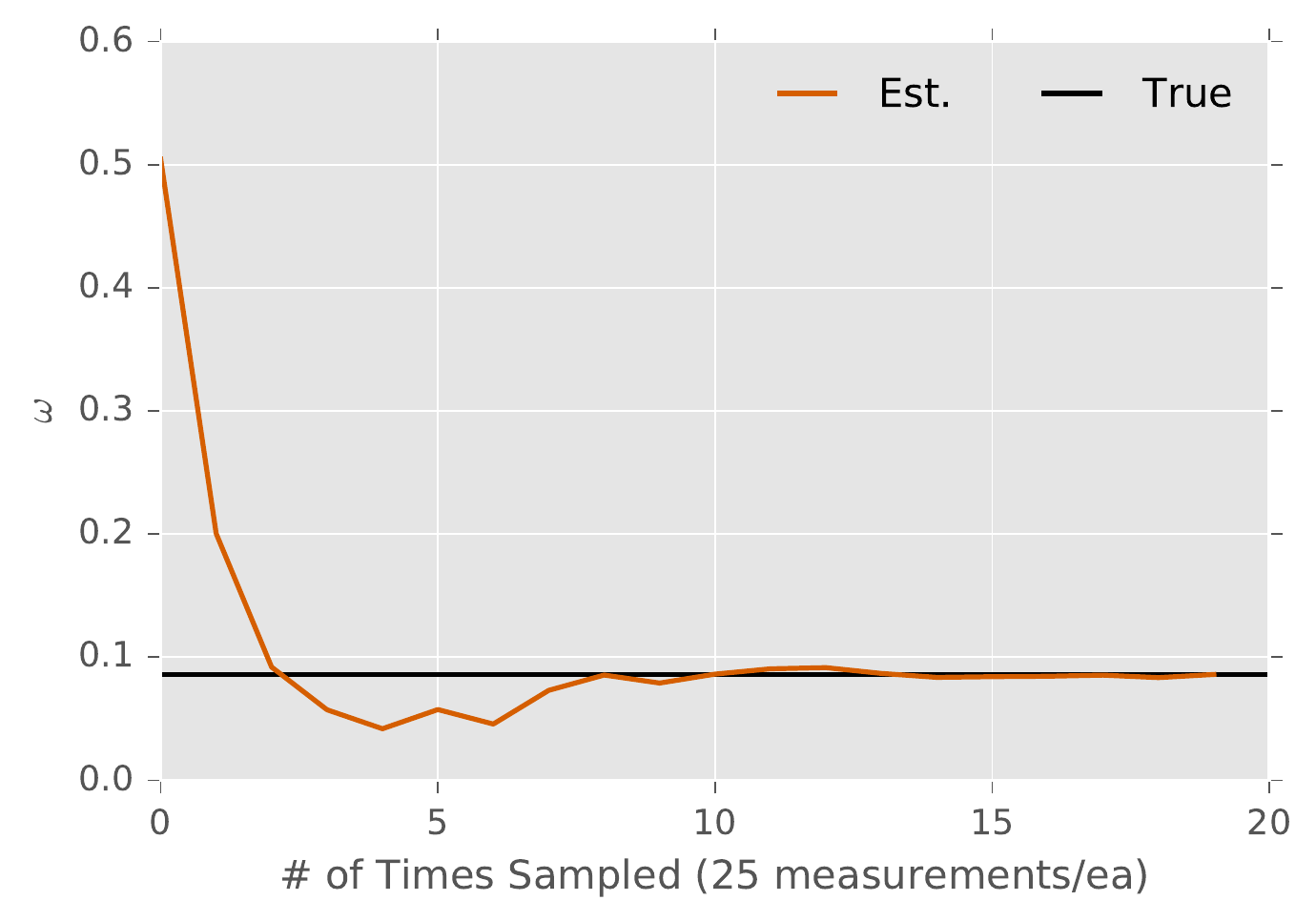}
        \caption{\label{fig:derived-model-updater-loop}
            Frequency estimate after 25 measurements at each
            of 20 linearly-spaced times, using \lstinline+qinfer.BinomialModel+
            as in \autoref{lst:derived-model-updater-loop}.
        }
    \end{center}
\end{figure}
\twocolumngrid

Note that parallelization, discussed in \autoref{sec:parallel}, is
implemented as a \lstinline+DerivedModel+ whose likelihood batches the
underlying model's likelihood function across processors.

\subsection{Time-Dependent Models}
\label{sec:time-dep}

So far, we have only considered time-independent (parameter estimation) models,
but particle filtering is useful more generally for estimating time-dependent
(state-space) models. Following the work of
\citet{isard_condensationconditional_1998}, when performing a Bayes update,
we may also incorporate state-space dynamics by adding a time-step update.
For example, to follow a Wiener process,
we move each particle $\vec{x}_i(t_k)$ at time $t_k$ to its
new position
\begin{equation}
    \label{eq:wiener-condensation}
    \vec{x}_i(t_{k+1}) = \vec{x}_t(t_k) + (t_{k+1} - t_k) \vec{\eta},
\end{equation}
with $\vec{\eta} \sim \N(0, \Sigma)$ for a covariance matrix $\Sigma$.

Importantly, we need not assume that time-dependence in $\vec{x}$
follows specifically a Wiener process. For instance, one
may consider timestep increments describing stochastic evolution of
a system undergoing weak measurement \cite{chase_single-shot_2009}, such
as an atomic ensemble undergoing probing by an optical interferometer
\cite{chase_magnetometry_2009}.
In each case, \qinfer~uses the timestep
increment implemented by the
\lstinline+Model.update_timestep+ method, which specifies the
time step that \lstinline+SMCUpdater+ should perform after each datum.
This design allows for the specification of more complicated time step updates
than the representative example of \autoref{eq:wiener-condensation}.
For instance, the \lstinline+qinfer.RandomWalkModel+ class adds diffusive
steps to existing models and can be used to quickly learn time-dependent
properties, such as shown in \autoref{lst:time-dep-rabi}.
Moreover, \qinfer~provides the \lstinline+DiffusiveTomographyModel+
for including time-dependence in tomography by truncating time step updates to
lie within the space of valid states \cite{granade_practical_2016}. A video
example of time-dependent tomography can be found on YouTube \cite{pbt-diffusion-video}.

In this way, by following the \citet{isard_condensationconditional_1998} algorithm, we
obtain a very general solution for time-dependence. Importantly, other approaches
exist that may be better suited for individual problems, including modifying
resampling procedures to incorporate additional noise
\cite{granade_characterization_2015,wiebe_efficient_2016}, or adding hyperparameters
to describe deterministic time-dependence \cite{granade_practical_2016}.

\onecolumngrid
\begin{lstlisting}[
        emph={RandomWalkModel},
        caption={Frequency estimation with a time-dependent
            model.},
        label={lst:time-dep-rabi}
    ]
    >>> from qinfer import *
    >>> import numpy as np
    >>> prior = UniformDistribution([0, 1])
    >>> true_params = np.array([[0.5]])
    >>> n_particles = 2000
    >>> model = RandomWalkModel(
    ...     BinomialModel(SimplePrecessionModel()), NormalDistribution(0, 0.01**2))
    >>> updater = SMCUpdater(model, n_particles, prior)
    >>> t = np.pi / 2
    >>> n_meas = 40
    >>> expparams = np.array([(t, n_meas)], dtype=model.expparams_dtype)
    >>> for idx in range(1000):
    ...     datum = model.simulate_experiment(true_params, expparams)
    ...     true_params = np.clip(model.update_timestep(true_params, expparams)[:, :, 0], 0, 1)
    ...     updater.update(datum, expparams)
\end{lstlisting}

\begin{figure}
    \begin{center}
        \includegraphics[width=0.7\columnwidth]{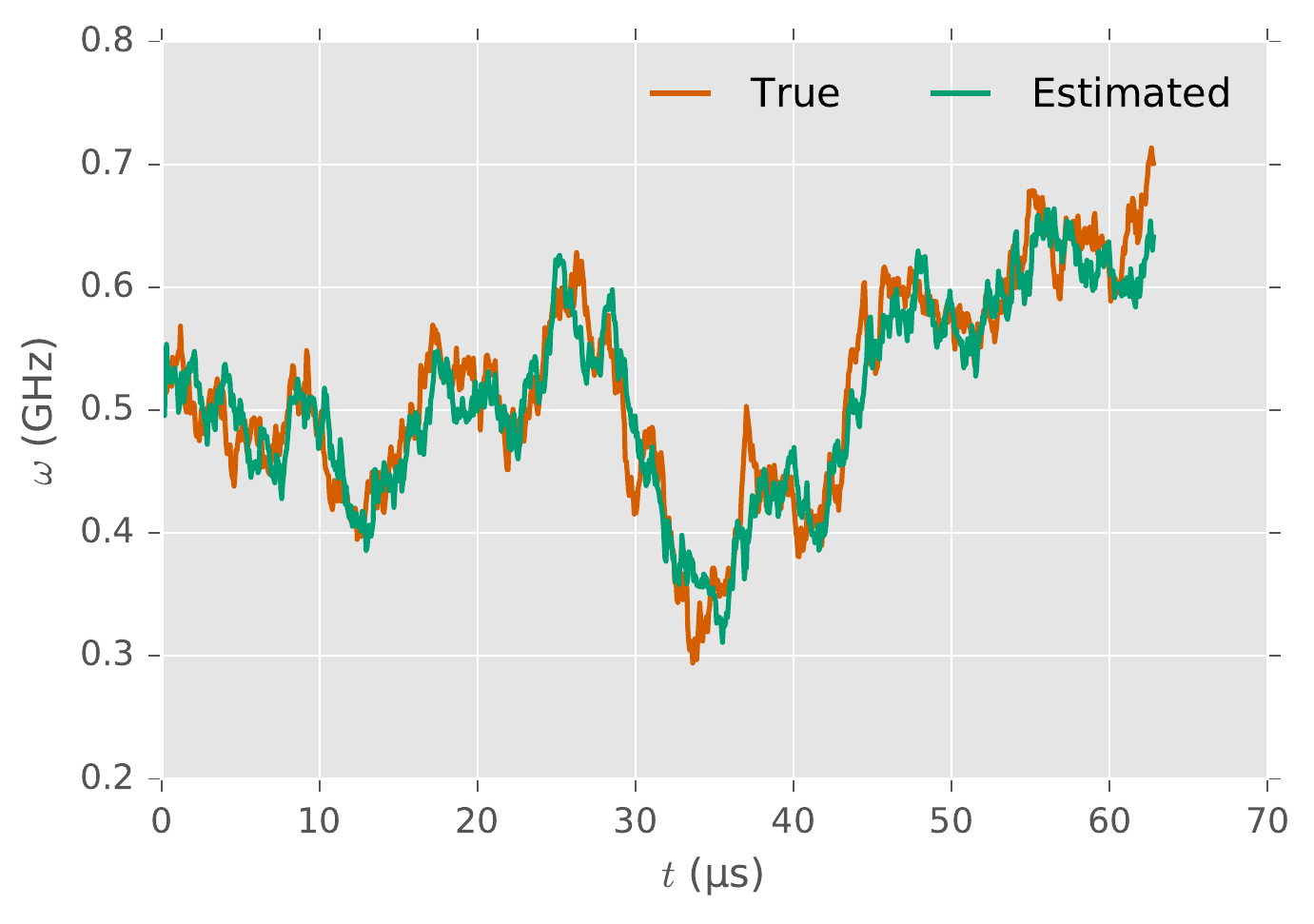}
        \caption{\label{fig:time-dep-rabi}
            Time-dependent frequency estimation,
            using \lstinline+qinfer.RandomWalkModel+
            as in \autoref{lst:time-dep-rabi}.
        }
    \end{center}
\end{figure}
\twocolumngrid

\subsection{Performance and Robustness Testing}
\label{sec:testing}

One important application of \qinfer~is predicting how well a particular
parameter estimation experiment will work in practice. This can be formalized
by considering the risk
$R(\vec{x}) \defeq \expect_D[(\hat{\vec{x}}(D) - \vec{x})^\T (\hat{\vec{x}}(D) - \vec{x})]$
incurred by the estimate $\hat{\vec{x}}(D)$ as a function of some true model
$\vec{x}$. The risk can be estimated by drawing many different data sets $D$,
computing the estimates for each, and reporting the average error. Similarly,
one can estimate the Bayes risk $r(\pi) \defeq \expect_{x \sim \pi}[R(\vec{x})]$
by drawing a new ``true'' model $\vec{x}$ from a prior $\pi$ along with
each data set.

In both cases, \qinfer~automates the process of performing many independent
estimation trials through the \lstinline+perf_test_multiple+ function.
This function will run an updater loop for a given model, prior,
and experiment design heuristic, returning the errors incurred after each
measurement in each trial. Taking an expectation value with \lstinline+numpy.mean+
returns the risk or Bayes risk, depending if the \lstinline+true_model+
keyword argument is set.

For example, \autoref{lst:bayes-risk} finds the
Bayes risk for a frequency estimation experiment (\autoref{sec:hamiltonian-learning})
as a function of the number of measurements performed.

Performance evaluation can also easily be parallelized over trials,
as discussed in \autoref{sec:parallel}, allowing for efficient use of
computational resources. This is especially important when comparing performance
for a range of different parameters. For instance, one might want to consider
how the risk and Bayes risk of an estimation procedure scale with errors in
a faulty simulator; \qinfer~supports this usecase with the \lstinline+qinfer.PoisonedModel+
derived model, which adds errors to an underlying ``valid'' model.
In this way, \qinfer~enables quickly reasoning about how much approximation
error can be tolerated by an estimation procedure.

\onecolumngrid
\begin{lstlisting}[
        caption={Bayes risk of frequency estimation as a function of the number of measurements,
            calculated using \lstinline+perf_test_multiple+\!.},
        mathescape=true,
        breaklines=true,
        emph={perf_test_multiple},
        label={lst:bayes-risk}
    ]
    >>> performance = perf_test_multiple(
    ...     # Use 100 trials to estimate expectation over data.
    ...     100,
    ...     # Use a simple precession model both to generate,
    ...     # data, and to perform estimation.
    ...     SimplePrecessionModel(),
    ...     # Use 2,000 particles and a uniform prior.
    ...     2000, UniformDistribution([0, 1]),
    ...     # Take 50 measurements with $t_k = ab^k$.
    ...     50, ExpSparseHeuristic
    ... )
    >>> # The returned performance data has an index for the trial, and an index for the measurement number.
    >>> print(performance.shape)
    (100, 50)
    >>> # Calculate the Bayes risk by taking a mean over the trial index.
    >>> risk = np.mean(performance['loss'], axis=0)
\end{lstlisting}

\begin{figure}
    \begin{center}
        \includegraphics[width=0.7\columnwidth]{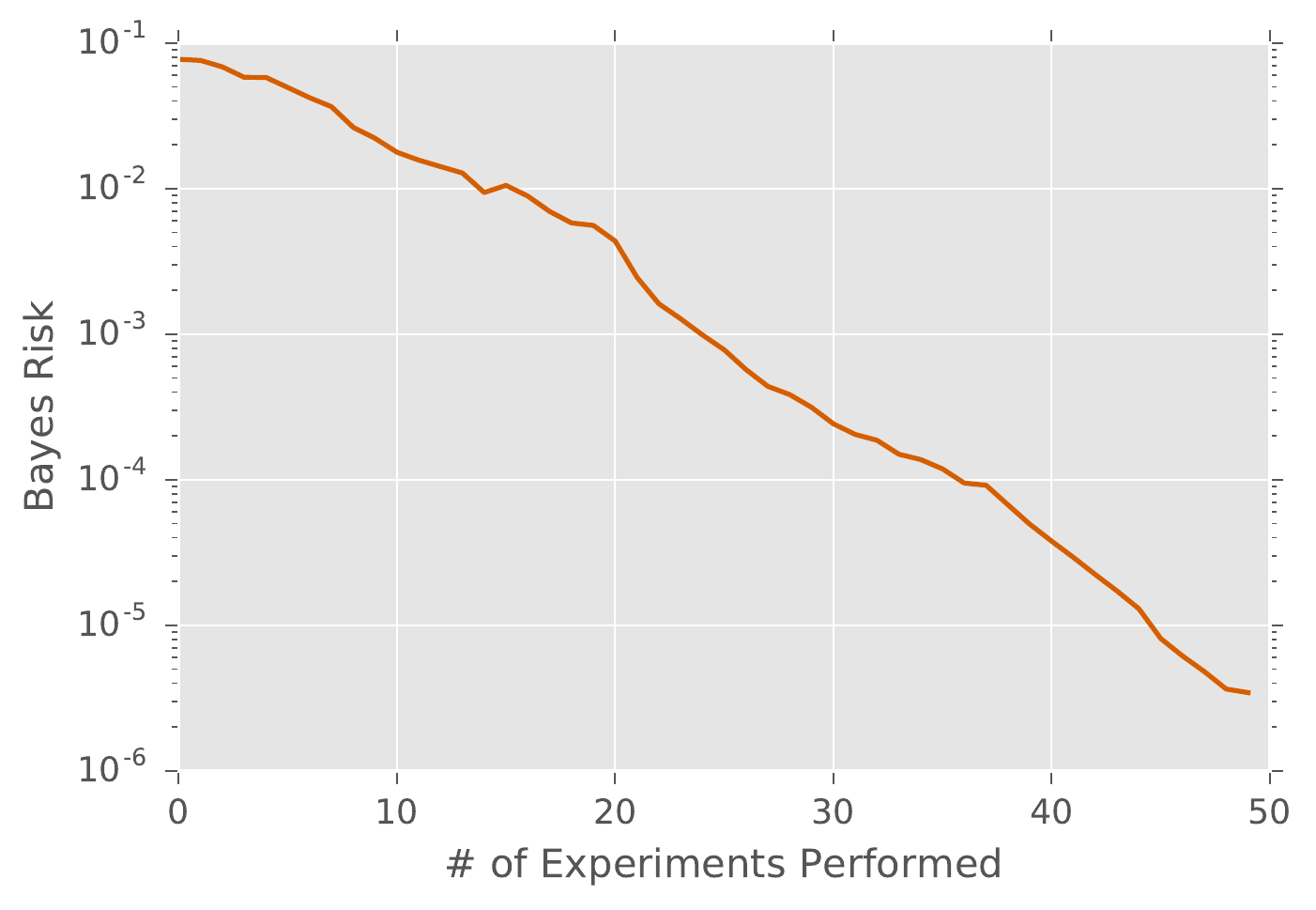}
        \caption{\label{fig:bayes-risk}
            Bayes risk of a frequency estimation model with exponentially sparse
            sampling as a function of the number of experiments performed,
            and as calculated by \autoref{lst:bayes-risk}.
        }
    \end{center}
\end{figure}
\twocolumngrid

\subsection{Parallelization}
\label{sec:parallel}

At each step of the SMC algorithm, the likelihood $\Pr(d_n|\vec{x})$
of an experimental datum $d_n$ is computed for every
particle $\vec{x}_k$ in the distribution.
Typically, the total running time of the algorithm is overwhelmingly
spent calculating these likelihoods.
However, individual likelihood computations are independent of each other
and therefore may be performed in parallel.
On a single computational node with multiple cores, limited parallelization
is performed automatically by relying on \numpy's vectorization primitives
\cite{walt_numpy_2011}.

More generally, however, if the running time of $\Pr(d_n|\vec{x})$ is largely
independent of $\vec{x}$, we may divide our particles into
$L$ disjoint groups,
\begin{align}
	\{\vec{x}_1^{(1)},...,\vec{x}_{k_1}^{(1)}\}
		\sqcup\cdots\sqcup
		\{\vec{x}_1^{(L)},...,\vec{x}_{k_L}^{(L)}\},
\end{align}
and send each group along with $d_n$ to a separate processor
to be computed in parallel.

In \qinfer, this is handled by the derived model (\autoref{sec:derived})
\lstinline+qinfer.DirectViewParallelizedModel+
which uses the Python library
\ipyparallel~\cite{ipython_development_team_ipyparallel_2016}.
This library supports everything from simple parallelization over
the cores of a single processor, to make-shift clusters
set up over SSH, to professional clusters using standard job schedulers.
Passing the model of interest as well as an \lstinline+ipyparallel.DirectView+
of the processing engines is all that is necessary to parallelize
a model.

\onecolumngrid
\begin{lstlisting}[
        emph={DirectViewParallelizedModel},
        caption={Example of parallelizing likelihood calls
            with \lstinline+DirectViewParallelizedModel+.},
        label={lst:parallel-ex},
        gobble=4
    ]
    >>> from qinfer import *
    >>> from ipyparallel import Client
    >>> rc = Client(profile="my_cores")
    >>> model = DirectViewParallelizedModel(SimplePrecessionModel(), rc[:])
\end{lstlisting}

\begin{figure}
	\begin{center}
		\includegraphics[width=0.7\columnwidth]{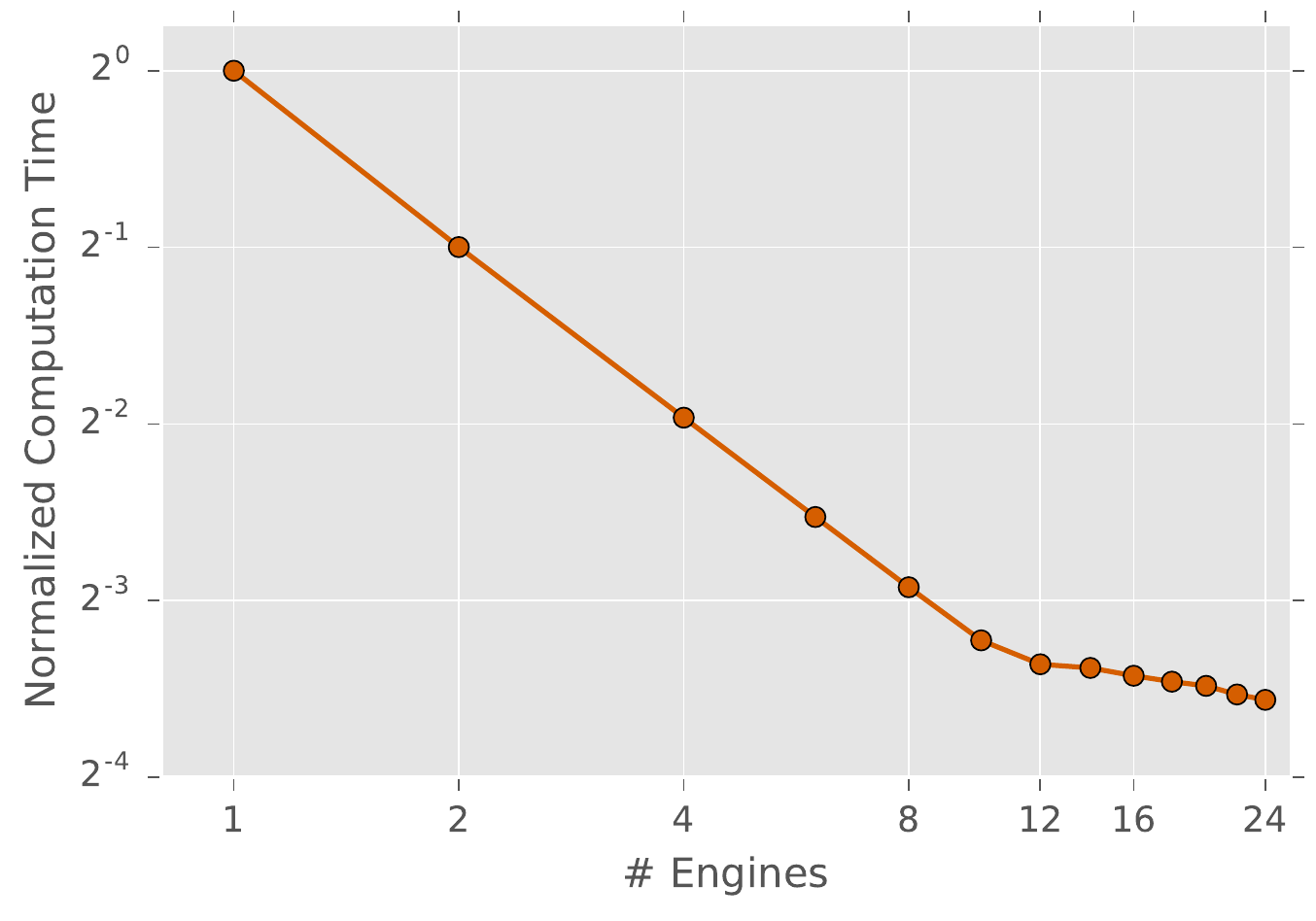}
        \caption{
            Parallelization of the likelihood function
            being tested on a single computer with 12 physical
            Intel Xeon cores. 5000 particles are shared over a varying number of
            \ipyparallel~engines.
            The linear unit slope indicates that overhead is negligible in this example.
            This holds until the number of physical cores is reached,
            past which hyper-threading continues to give diminishing returns.
            The single-engine running time was about 37 seconds, including
            ten different experiment values, and 5 possible outcomes.
        }
        \label{fig:parallel-likelihood}
    \end{center}
\end{figure}
\twocolumngrid

In \autoref{fig:parallel-likelihood}, a roughly 12$\times$ speed-up is
demonstrated by parallelizing a model over the 12 cores of a single
computer.
This model was contrived to demonstrate the parallelization
potential of a generic Hamiltonian learning problem which uses
dense operators and states.
A single likelihood call generates a random $16\times 16$ anti-hermitian
matrix (representing the generator of a four qubit system),
exponentiates it, and returns overlap with the $\ket{0000}$ state.
Implementation details can be found in the \qinfer~examples repository
\cite{qinfer-examples}, or in the ancillary files.

So far, we have discussed parallelization from the perspective of traditional
processors (CPUs), which typically have a small number of processing cores on
each chip. By contrast, moderately-priced desktop
graphical processing units (GPUs) will often contain thousands of cores, while
GPU hosts tailored for scientific use can have tens of thousands.
This massive parallelization makes GPUs attractive for particle filtering
\cite{lee_utility_2010}.
Using libraries such as PyOpenCL and PyCUDA \cite{kloeckner_pycuda_2012}
or Numba \cite{lam_numba:_2015}, custom models can be written which take advantage of GPUs within
\qinfer~\cite{granade_characterization_2015}. For example,
\lstinline+qinfer.AcceleratedPrecessionModel+ offloads its computation of
$\cos^2$ to GPUs using PyOpenCL.

\subsection{Other Features}

In addition to the functionality described above, \qinfer~has a wide range
of other features that we describe more briefly here. A complete description
can be found in the provided Users' Guide (see ancillary files or
\url{docs.qinfer.org}).

\paragraph{Plotting and Visualization}

\qinfer~provides plotting and visualization support based on matplotlib
\cite{Hunter:2007} and mpltools \cite{yu_mpltools_2015}. In particular,
\lstinline+qinfer.SMCUpdater+ provides methods for plotting posterior distributions
and covariance matrices. These methods make it straightforward to visually
diagnose the operation of and results obtained from particle filtering.

Similarly, the \lstinline+qinfer.tomography+ module provides several functions
for producing plots of states and distributions over rebits (qubits restricted
to real numbers). Rebit visualization is in particular useful for demonstrating
the conceptual operation of particle filter--based tomography in a clear and
attractive manner.

\paragraph{Fisher Information Calculation}

In evaluating estimation protocols, it is important to establish a baseline of
how accurately one can estimate a model even in principle. Similarly, such a
baseline can be used to compare between protocols by informing as to how much
information can be extracted from a proposed experiment. The Cram\'er--Rao
bound and its Bayesian analog, the van Trees inequality (\emph{a.k.a.} the
Bayesian Cram\'er--Rao bound), formalize this notion in terms of the Fisher
information matrix \cite{cover_elements_2006,gill_applications_1995}. For any
model which specifies its derivative in terms of a \emph{score}, \qinfer~will
calculate each of these bounds, providing useful information about proposed
experimental and estimation protocols. The \lstinline+qinfer.ScoreMixin+ class
builds on this by calculating the score of an arbitrary model using numerical
differentiation.

\paragraph{Model Selection and Averaging}

Statistical inference does not require asserting \emph{a priori} the
correctness of a particular model (that is, likelihood function), but allows
a model to be taken as a hypothesis and compared to other models. This is made
formal by \emph{model selection}. From a Bayesian perspective, the ratio
of the posterior normalizations for two different models gives a natural
and principled model selection criterion, known as the Bayes factor
\cite{jeffreys_theory_1998}.
The Bayes factor provides a model selection rule that is significantly
more robust to outlying data than conventional hypothesis testing
approaches \cite{edwards_bayesian_1963}.
For example, in quantum applications, the Bayes factor
is particularly useful in tomography, and can be used
to decide the rank or dimension of a state \cite{ferrie_quantum_2014}.
\qinfer~implements this criterion as the \lstinline+SMCUpdater.normalization_record+
property, allowing for model selection and averaging to be performed in a
straightforward manner.

\paragraph{Approximate Maximum-Likelihood Estimation}

As opposed to the Bayesian approach, one may also consider \emph{maximum
likelihood} estimation (MLE), in which a model is estimated as $\hat{\vec{x}}_{\text{MLE}}
\defeq \operatorname{arg\ max}_{\vec{x}} \Pr(D | \vec{x})$. MLE can be approximated
as the mean of an artificially tight posterior distribution obtained by performing
Bayesian inference with a likelihood function ${\Pr}'(D | \vec{x})$ related to the
true likelihood by
\begin{equation}
    {\Pr}'(D | \vec{x}) = \left(\Pr(D | \vec{x})\right)^\gamma
\end{equation}
for a quality parameter $\gamma > 1$ \cite{johansen_particle_2008}.
Similarly, taking $\gamma < 1$ with appropriate
resampling parameters allows the user to \emph{anneal} updates \cite{deutscher_articulated_2000},
avoiding the dangers posed by strongly multimodal likelihood functions.
In this way, taking $\gamma < 1$ is roughly
analogous to the use of ``reset rule'' techniques employed in other
filtering algorithms \cite{wiebe_efficient_2016}.
In \qinfer, both cases are implemented by the class \lstinline+qinfer.MLEModel+,
which decorates another model in the manner of \autoref{sec:derived}.

\paragraph{Likelihood-Free Estimation}

For some models, explicitly calculating the likelihood function $\Pr(D | \vec{x})$
is intractable, but good approaches may exist for drawing new data sets
consistent with a hypothesis. This is the case, for instance, if a quantum
simulator is used in place of a classical algorithm, as recently proposed for
learning in large quantum systems \cite{wiebe_hamiltonian_2014}.
In the absence of an explicit likelihood
function, Bayesian inference must be implemented in a \emph{likelihood-free}
manner, using hypothetical data sets consistent to form a approximate likelihood
instead \cite{ferrie_likelihood-free_2014}. This introduces an estimation error
which can be modeled in \qinfer~by using the \lstinline+qinfer.PoisonedModel+
class discussed in \autoref{sec:testing}.

\paragraph{Simplified Estimation}

For the frequency estimation and randomized benchmarking examples described
in \autoref{sec:quantum-models}, \qinfer~provides functions to perform estimation
using a ``standard'' updater loop, making it easy to load data from \numpy-,
MATLAB- or CSV-formatted files.

\paragraph{Jupyter Integration}

Several \qinfer~classes, including \lstinline+qinfer.Model+ and
\lstinline+qinfer.SMCUpdater+, integrate with Jupyter Notebook to provide
additional information formatted using HTML. Moreover, the
\lstinline+qinfer.IPythonProgressBar+ class provides a progress bar
as a Jupyter Notebook widget with a \qutip-compatible interface, making it
easy to report on performance testing progress.

\paragraph{MATLAB/Julia Interoperability}

Finally, \qinfer~functionality is also compatible with MATLAB 2016a and later,
and with Julia (using the \lstinline+PyCall.jl+ package \cite{johnson_pycall.jl_2016}),
enabling integration both with legacy code and with new developments in scientific
computing.

\section{Conclusions}
\label{sec:conclusions}

In this work, we have presented \qinfer, our open-source library for
statistical inference in quantum information processing. \qinfer~is useful for
a range of different applications, and can be readily used for custom problems
due to its modular and extensible design, addressing a pressing need in both
quantum information theory and in experimental practice. Importantly, our
library is also accessible, in part due to the extensive documentation that we
provide (see ancillary files or \url{docs.qinfer.org}). In this way,
\qinfer~supports the goal of reproducible research by providing open-source
tools for data analysis in a clear and understandable manner.


\acknowledgments{
    CG and CF acknowledge funding from the Army Research Office
    grant numbers W911NF-14-1-0098 and W911NF-14-1-0103, from the
    Australian Research Council Centre of Excellence for Engineered Quantum Systems.
    CG, CF, IH, and TA acknowledge funding from Canadian Excellence Research Chairs (CERC),
    Natural Sciences and Engineering Research Council of Canada (NSERC),
    the Province of Ontario, and Industry Canada.
    JG acknowledges funding from ONR Grant No. N00014-15-1-2167.
    YS acknowledges funding from ARC Discovery Project DP160102426.
    CG greatly appreciates help in testing and feedback
    from Nathan Wiebe, Joshua Combes, Alan Robertson, and Sarah Kaiser.
}

\nocite{apsrev41Control}
\bibliographystyle{apsrev4-1}
\bibliography{apsrev-control,qinfer,supplemental-material}


\appendix
\onecolumngrid

\section{Custom Model Example}
\label{apx:custom-model}

In \autoref{lst:multicos-model}, below, we provide an example of a
custom subclass of \lstinline+qinfer.FiniteOutcomeModel+ that
implements the likelihood function
\begin{align}
    \label{eq:multicos-like}
    \Pr(0 | \omega_1, \omega_2; t_1, t_2) & = \cos^2(\omega_1 t_1 / 2) \cos^2(\omega_2 t_2 / 2)
\end{align}
for model parameters $\vec{x} = (\omega_1, \omega_2)$ and experiment parameters
$\vec{e} = (t_1, t_2)$. A more efficient implementation of this model using
\numpy~vectorization is presented in more detail in the Users' Guide.

\begin{lstlisting}[
        gobble=4,
        label={lst:multicos-model},
        caption={
            Example of a custom \lstinline+FiniteOutcomeModel+ subclass implementing
            the multi-$\cos$ likelihood \autoref{eq:multicos-like}.
        },
        emph={FiniteOutcomeModel}
    ]
    from qinfer import FiniteOutcomeModel
    import numpy as np

    class MultiCosModel(FiniteOutcomeModel):

        @property
        def n_modelparams(self):
            return 2

        @property
        def is_n_outcomes_constant(self):
            return True

        def n_outcomes(self, expparams):
            return 2

        def are_models_valid(self, modelparams):
            return np.all(np.logical_and(modelparams > 0, modelparams <= 1), axis=1)

        @property
        def expparams_dtype(self):
            return [('ts', 'float', 2)]

        def likelihood(self, outcomes, modelparams, expparams):
            super(MultiCosModel, self).likelihood(outcomes, modelparams, expparams)
            pr0 = np.empty((modelparams.shape[0], expparams.shape[0]))

            w1, w2 = modelparams.T
            t1, t2 = expparams['ts'].T

            for idx_model in range(modelparams.shape[0]):
                for idx_experiment in range(expparams.shape[0]):
                    pr0[idx_model, idx_experiment] = (
                        np.cos(w1[idx_model] * t1[idx_experiment] / 2) *
                        np.cos(w2[idx_model] * t2[idx_experiment] / 2)
                    ) ** 2

            return FiniteOutcomeModel.pr0_to_likelihood_array(outcomes, pr0)
\end{lstlisting}

\section{Resampling}
\label{apx:resampling}

The purpose of this appendix is to offer a brief discussion of resampling
for particle filters.  In \qinfer, the standard resampler is the one
proposed by Liu and West~\cite{liu_combined_2001}.  We begin by motivating
the development of resamplers by explaining the problem of impoverishment
in a particle filter.  We then describe resamplers by explaining that their
effect is to produce a particle filter that approximates a smoothed version
of the underlying probability distribution.  Finally, we explain the Liu--West
resampling algorithm.

Particle filters are intended to allow for the approximation of
the expectation values of functions, but admit an ambiguity between
using the locations and the weights to do so.  Assuming that the particle locations
are primarily responsible for representing the particle filtering
approximation, and assuming that we want to approximate the expectation value
of a function that is not pathological in some way, the number of particles
then serves as a reasonable proxy for the quality of the particle filter.
This follows from exactly the same argument as for Monte Carlo integration,
as in this case, the particle locations can be seen to directly correspond to
samples of an integrand.
On the other hand, if either of these assumptions are violated, one cannot trust
the numerical answers obtained using the particle filter method without
an additional argument. Since the weights will in general become less even as
Bayesian inference proceeds, we will rely on resampling to provide us with
precisely such an argument.

More precisely, the purpose of resampling is to mitigate against the loss in numerical
stability caused by having a large number of low-weight particles.
If a particle has a small weight, we could neglect it from the computation
of an expectation value without introducing a large difference in the result,
such that it no longer contributes to the approximation quality of \autoref{eq:smc-approx}.
That is, the particle's effectiveness at contributing to the stability
of the algorithm decreases as its weight decreases.
This observation then motivates using the \emph{effective sample size}
\begin{equation}
    n_\ess := \frac{1}{\sum_k w_k^2}
\end{equation}
as a criterion to ensure the numerical stability of particle filtering
\cite{beskos_stability_2014}.

In the case of roughly equal weights, the effective sample size
$n_\ess$ is roughly equal to the actual number of particles $n$.
If the weights are distributed rather unevenly, however, $n_\ess \ll n$.
In the latter case, we say that the particle filter is \emph{impoverished}.
Notably, if a particle filter becomes extremely impoverished, it may not be feasible
to effectively recover numerical stability, such that the minimum observed
value of $n_\ess$ serves as a diagnostic criteria for when the numerical
approximations used by particle filtering have failed \cite{granade_paqt_2016}.
For this reason, \qinfer~will by default call the resampler when $n_\ess$
falls below half of its initial value, and will warn if $n_\ess \le 10$
is ever observed.

Impoverishment is the result of choosing particle locations according to prior
information alone; with near certainty, the posterior distribution will be tightly
centered away from all initial particle locations such that some re-discretization
will be needed to represent the final posterior.
The goal of resampling is therefore to modify the choice of particle locations
not based on the interpretation of new data, but rather to ensure that the particles
are concentrated so as to accurately represent the probability distribution
of interest.
A resampling algorithm is then any algorithm designed to replace
an input particle filter that may be impoverished with a new particle filter
that is not impoverished but approximates the same probability distribution.
Since such a procedure cannot exist in full generality, each resampling procedure
works by enforcing that a particular set of invariants remains true before and after
the resampling step.

Returning to the particular case of the Liu--West resampler~\cite{liu_combined_2001}
employed by default in \qinfer,
we note that the Liu--West algorithm is particularly simple to understand
from the perspective of kernel density estimation \cite{west_approximating_1993}.
We will not provide a complete algorithm here; such explicit algorithms
can be found many places in the literature, including
Algorithm~2 of~\citet{granade_characterization_2015} and
Algorithm~2.5 of~\citet{sanders_thesis}. Rather, we explain that
the Liu--West algorithm acts by preserving the first two moments (the
expectation and the variance) of its input distribution.

The Liu--West algorithm starts by using that, as in the celebrated
method of kernel density estimation, each sample from a continuous distribution can be
used to infer properties about the neighborhood around that sample, given moderate
assumptions on the smoothness of a distribution.
Thus, the particle filter might be used to define a direct
approximation to the true probability distribution by first defining some function $K$,
called a \emph{kernel}, such that the estimated distribution
\begin{equation}
    \sum_k w_k K \left( \vec{x} - \vec{x}_k \right)
\end{equation}
is a good approximation of the true probability distribution.
This then leaves open the question of what kernel function $K$ should be chosen.
Owing to its generality, choosing $K$ to be a normal distribution works well under
only mild assumptions, leaving us to choose the variance of the kernel.
Specializing to the single-parameter case for simplicity of notation,
we denote Gaussian kernels as $K(x; \sigma^2)$, where $\sigma^2$ is the
variance to be chosen. The multi-parameter case follows by replacing this variance
by a covariance matrix.

The key insight at the core of the Liu--West algorithm is the observation that
the posterior distribution will narrow over time, so that we can choose the
variance of the kernel to narrow over time in proportion to the variance of
the distribution being approximated. In particular, let $a \in [0, 1]$ be a
parameter and let $h$ such that $a^2 + h^2 = 1$. Then choose the kernel to be
$K(x; h^2 \mathbb{V}[x])$. Since this alone would increase the variance
of the distribution under approximation to $1 + h^2$, the Liu--West resampler
instead draws new particle locations $x'$ from the distribution
\begin{align}
    x' \sim \sum_k w_k K(x - (a x_k + (1 - a) \expect[x]; h^2 \mathbb{V}[x]).
\end{align}
This distribution contracts each original particle towards the mean by $1 - a$,
such that the mean and variance of the post-resampling distribution are identical
to the distribution being approximated.

Importantly, the Liu--West generalizes existing resampling algorithms, such that
the bootstrap \cite{doucet_tutorial_2011} and assumed density filtering resamplers
\cite{minka_family_2001,wiebe_efficient_2016} are given by $a = 1$ and $a = 0$,
respectively. We also note that violating the invariant that $\mathbb{V}[x]$ is preserved
can allow for some robustness to multimodal distributions and time-dependence
\cite{granade_characterization_2015}. Finally, a video of the Liu--West resampler
applied to Bayesian inference on the model
of \autoref{lst:multicos-model} is available online \cite{multicos-resampling-video}.

\end{document}